# Lieb polariton topological insulators

Chunyan Li,[1] Fangwei Ye,[1,*] Xianfeng Chen,[1] Yaroslav V. Kartashov,[2,3,4] Albert Ferrando,[5] Lluis Torner,[2,6] and Dmitry V. Skryabin[4,7]

[1]*Key Laboratory for Laser Plasma (Ministry of Education), Collaborative Innovation Center of IFSA (CICIFSA), Department of Physics and Astronomy, Shanghai Jiao Tong University, Shanghai 200240, China.*

[2]*ICFO-Institut de Ciencies Fotoniques, The Barcelona Institute of Science and Technology, 08860 Castelldefels (Barcelona), Spain*

[3]*Institute of Spectroscopy, Russian Academy of Sciences, Troitsk, Moscow, 142190, Russia*

[4]*Department of Physics, University of Bath, BA2 7AY Bath, UK*

[5]*Departament d'Òptica, Interdisciplinary Modeling Group, InterTech, Universitat de València, València, Spain*

[6]*Universitat Politecnica de Catalunya, 08034, Barcelona, Spain*

[7]*ITMO University, St. Petersburg 197101, Russia*

*\*Corresponding author: fangweiye@sjtu.edu.cn*

We predict that the interplay between the spin-orbit coupling, stemming from the TE-TM energy splitting, and the Zeeman effect in semiconductor microcavities supporting exciton-polariton quasi-particles results in the appearance of unidirectional linear topological edge states when the top microcavity mirror is patterned to form a truncated dislocated

Lieb lattice of cylindrical pillars. Periodic nonlinear edge states are found to emerge from the linear ones. They are strongly localized across the interface and they are remarkably robust in comparison to their counterparts in hexagonal lattices. Such robustness makes possible the existence of nested unidirectional dark solitons that move steadily along the lattice edge.

Topological insulation is a recently discovered fundamental phenomenon that spans across several areas of physics, such as condensed matter, ultracold gases, and photonics [1,2]. In contrast to conventional insulators, topological ones admit conductance at the interfaces between materials with different topologies. This conductance is a consequence of the existence of the in-gap discrete energy states spatially localized at the boundaries and exhibiting unidirectional propagation. The latter is topologically protected and hence stays immune to the backward scattering and energy leakage to the bulk insulator modes even when encountered with strong lattice defects and disorder [1,2]. First experiments about topological insulators were performed in electronic systems where one of the mechanisms of topological protection relies on the spin-orbit interaction of electrons in a magnetic field [1,2]. More recently, studies on topological edge effects have been extended to electromagnetic and to mixed opto-electronic systems [3]. Topological edge states have been proposed and observed in gyromagnetic photonic crystals [4,5], semiconductor quantum wells [6], arrays of coupled resonators [7,8], metamaterial superlattices [9], helical photonic waveguide arrays [10,11], systems with driving fields containing vortex lattices [12], and in polariton microcavities, where strong photon-exciton coupling leads to formation of the half-light half-matter polariton quasi-particles [13-17].

Matter-wave and polaritonic systems are especially attractive in the context of topological photonics because they are strongly nonlinear [18-26] and therefore potentially applicable in classical and quantum information processing schemes. Examples of recently reported nonlinear topological effects include solitons in bulk [27] and at the edges [28-30] of topological insulators made from the hexagonal lattices of helical photonic waveguides. We also mention here prior reports of the non-topological nonlinear edge states in photonic lattices [31,32]. Microcavity exciton-polaritons represent a viable alternative to photons as a nonlinear platform for topological effects [13,33-35]. Polaritonic systems are planar and rely on the interplay between the spin-orbit coupling (SOC) with the Zeeman energy splitting of polariton energy levels induced by a magnetic field. They have been shown to support long-living topological solitons in honeycomb [36] and Kagome lattice potentials [37].

Recently, Lieb lattice potentials have attracted significant attention due to their flat energy bands [38-42] associated with the infinite mass bosons. This implies that dispersion and kinetic energy are suppressed [40] and therefore the system dynamics is dominated by the multi-particle interaction, i.e., by the nonlinear effects. Bosonic condensation, fractional Hall effects, and the topological insulator regimes in the flat bands and more generally in the Lieb lattices can be qualitatively different from those known for quasi-particles with a well-defined effective mass [38,39,41,43,44].

In this work, we show that robust nonlinear topological edge states exist in the polariton excitations embedded into *dislocated* Lieb lattice potentials (a variant of Lieb lattice obtained by vertical displacement of its adjacent unit cells by half-period [45,46] that admits substantially larger topological gaps than usual Lieb lattice), in the presence of SOC and Zeeman splitting. We show that the nonlinear properties of the Lieb polariton topological insulators are dominated by the existence of robust, topological dark edge solitons.

We model the evolution of the spinor polariton wavefunction $\mathbf{\Psi}=(\psi_+,\psi_-)^\mathrm{T}$ in a lattice of microcavity pillars by the system of two coupled Gross-Pitaevskii equations [13]:

$$i\frac{\partial \psi_\pm}{\partial t}=-\frac{1}{2}\left(\frac{\partial^2}{\partial x^2}+\frac{\partial^2}{\partial y^2}\right)\psi_\pm+\beta\left(\frac{\partial}{\partial x}\mp i\frac{\partial}{\partial y}\right)^2\psi_\mp+R(x,y)\psi_\pm\pm\Omega\psi_\pm+(|\psi_\pm|^2+\sigma|\psi_\mp|^2)\psi_\pm \quad (1)$$

Here $\psi_\pm$ are the spin-positive and spin-negative wave function components in the circular polarization basis. They are related to the wave functions corresponding to the TE (subscript $y$) and TM (subscript $x$) polarizations as $\psi_\pm=(\psi_\mathrm{x}\mp i\psi_\mathrm{y})/2^{1/2}$. The spin-orbit coupling term $\sim\beta$ originates in the TE-TM energy splitting of the cavity resonances in the lattice-free environment [47]. $\Omega$ is the Zeeman splitting in the external magnetic field. Polaritons with the same spins repel and with the opposite ones attract. The latter implies that the parameter $\sigma$ of the cross-spin interaction is negative. We fix $\sigma=-0.05$ [22]. The potential landscape $R(x,y)=-p\sum_{km}e^{-[(x-x_k)^2+(y-y_m)^2]/d^2}$ is created by the microcavity pillars (the contribution from each pillar is described by a Gaussian function of width $d$ and depth $p$) arranged into the *dislocated* Lieb lattice. An example of such a lattice with three $y$-periods is shown in Fig. 1(a). The lattice is infinite along $y$-axis and is truncated along $x$ (we consider truncation creating interface that can be named *bearded* by analogy with honeycomb lattice [10]).

In Eqs. (1) we assume that all the distances are scaled to $x_0 = 1\,\mu\mathrm{m}$, all the energy parameters to $\varepsilon_0 = \hbar^2/2mx_0^2$, and the time to $t_0 = \hbar\varepsilon_0^{-1}$. By selecting the polariton mass parameter $m = 10^{-31}\,\mathrm{g}$ one gets the characteristic energy $\varepsilon_0 \approx 0.35\,\mathrm{meV}$ and time $t_0 \approx 1.9\,\mathrm{ps}$. The depth of the potential $p = 8$ corresponds to $\sim 2.8\,\mathrm{meV}$ and the width $d = 0.5$ of each potential well corresponds to $0.5\,\mu\mathrm{m}$. If taken in isolation, such pillars support only the ground state mode. We set the spacing between micropillars to be $a = 1.4$, corresponding to $1.4\,\mu\mathrm{m}$. Since the existence of the topological edge states is not connected with the presence of losses, here we focus on the quasi-conservative limit, which has been previously studied both experimentally [19,23,24,34] and theoretically [13,14,47].

First we consider the spectrum of the linear Bloch modes $\psi_\pm(x,y,t) = u_\pm(x,y)e^{iky-i\varepsilon t}$ that are periodic along the $y$-axis [$u_\pm(x,y) = u_\pm(x,y+2a)$] and localized along the $x$-axis [$u_\pm(x\to\pm\infty,y) = 0$]. Here $k$ is the Bloch momentum and $\varepsilon$ is the energy shift relative to the bottom of the polariton energy-momentum characteristic. A unit cell used for the calculation of the Bloch modes contains 21 periods along $x$ and 1 along the $y$ direction [Fig. 1(a)]. Typical spectra $\varepsilon(k)$ are shown in Fig. 2 for $k \in [0,\mathrm{K}]$, where $\mathrm{K} = \pi/a$ is the Brillouin zone width. Due to the spinor nature of our problem, the spectrum consists of the two families of bands, which are degenerate if SOC and Zeeman splitting are disregarded: $\Omega = 0$, $\beta = 0$. Accounting for the Zeeman splitting $\Omega = 0.5$ leads to a relative shift of the two families by $\delta\varepsilon = 2\Omega$, as shown in Fig. 2(a). When $\beta = 0$ the system does not admit topo-logical edge states [Fig. 2(a)]. Inclusion of SOC into the system ($\beta \neq 0$) at $\Omega \neq 0$ breaks the time-reversal symmetry of Eqs. (1). Following Ref. [4], this should lead to the *opening of topological gaps* around special points in the Lieb lattice spectrum, and, if the lattice is truncated, to the appearance of topological in-gap edge states branching off the boundaries of the bulk bands. Topological gaps open wider with the increase of SOC. Topological edge states found for $\beta = 0.3$ are shown as red, green, blue, and magenta curves in Fig. 2(b), while the black curves correspond to the bulk modes. The states with the same energy $\varepsilon$, but with different momenta $k < K/2$ and $k > K/2$ reside on the opposite edges of the lattice [cf. second and fifth rows in Fig. 1] and have opposite group velocities $\varepsilon' = \partial \varepsilon / \partial k$. This is a direct manifestation of *unidirectional edge transport* in the Lieb polariton topological insulator. Thus, states from blue and green branches reside on the left edge, while states from red and magenta branches reside on the right edge [see examples in Fig. 1]. Edge localization is most pronounced when the topological gap state has energy close to the centre of the gap and decreases when it approaches the band. All the edge states in Fig. 2 feature dominating $\psi_-$ and weak $\psi_+$ components – a consequence of the selected sign of magnetic field. First-order $\varepsilon' = \partial \varepsilon / \partial k$ and second-order $\varepsilon'' = \partial^2 \varepsilon / \partial k^2$ dispersion coefficients associated with blue and magenta topological branches are shown in Fig. 3(a). While $\varepsilon'$ can change its sign upon variation of $k$, the $\varepsilon''$ coefficient remains positive for selected branches.

To confirm that the above edge states are indeed topological we calculated the Chern numbers associated with the infinite lattice using the algorithm described in [48]. Chern number for the nth band is given by $C_n = (2\pi i)^{-1} \int_G F(\mathbf{k}) d^2\mathbf{k}$, where G denotes the first

Brillouin zone and the function $F(\mathbf{k})=\partial_x A_y(\mathbf{k})-\partial_y A_x(\mathbf{k})$ can be determined from the Berry connection $A_{x,y}(\mathbf{k})=\langle\psi(\mathbf{k})|\partial/\partial k_{x,y}|\psi(\mathbf{k})\rangle$. Our calculations confirmed that for $\Omega=0.5$ at $\beta=0$, i.e. without SOC, the Chern numbers for all bands are zero. This is consistent with the absence of unidirectional topological edge states at $\beta=0$. However, when $\beta\neq0$ at $\Omega=0.5$, the calculated Chern numbers for the first three bands (from the bottom to top) are -1, 0 and +1 respectively. This means that the *gap* Chern numbers for the first three gaps are -1,-1 and 0, a direct indication of the appearance of one topological edge mode (per interface) in the first and the second gap, and no edge modes in the third gap. Note that these results are consistent with the modal analysis for the truncated dislocated Lieb lattice discussed above (our lattice is truncated on both sides, so that the number of edge states in each gap is twice the respective gap Chern number). Moreover, to show that the inclusion of SOC transforms non-topological modes into topological ones, we depict in Fig. 4 the variation of the profile of the mode from green branch at $k=0.40\mathrm{K}$ with the increase of $\beta$. At $\beta=0$ the distribution of $|\psi_-|$ is symmetric in $x$, the field is nonzero at both interfaces. Increasing $\beta$ leads to the appearance of the topological edge state, whose localization increases with $\beta$.

As a third indication of the topological nature of the linear edge states we studied their evolution in a lattice where one pillar at the lattice edge is missing. The propagation of the edge state with a Gaussian envelope in such lattice with a defect is displayed in Fig. 5. The wavepacket smoothly bypasses defect without noticeable backscattering or radiation into the bulk. The topological protection for these edge modes is thus evident.

Having elucidated the existence of unidirectional edge states in Lieb polariton insulators, we now focus on nonlinear edge states. They are sought in the form $\psi_\pm(x,y,t)=u_\pm(x,y)e^{iky-i\mu t}$, where $\mu$ is the nonlinearity-induced energy shift ($\mu$ becomes $\varepsilon$ in the linear limit), and $u_\pm(x,y)$ is periodic in $y$ with period $2a$. The nonlinear edge states are expected to exist for $\mu(k) \geq \varepsilon(k)$. We computed them numerically using the Newton method in the Fourier domain and characterized them using the peak amplitudes of the spin-positive $a_+ = \max|\psi_+|$ and the spin-negative $a_- = \max|\psi_-|$ components, and the norm per $y$-period

$$U = \int_0^{2a} dy \int_{-\infty}^{+\infty} (|\psi_+|^2 + |\psi_-|^2) dx. \tag{2}$$

These parameters are plotted in Fig. 3(b) as functions of $\mu$ for the edge state bifurcating from the blue linear branch of Fig. 2(b) at $k=0.2\,\mathrm{K}$. The nonlinear edge states appear to be thresholdless, since $a_\pm$ vanish in the bifurcation point. One has $a_- > a_+$ for the nonlinear states, a property inherited from the linear limit. The norm and the peak amplitudes monotonically increase with $\mu$ until the edge of the second linear band is reached for a given $k$ (dashed line). When $\mu$ crosses the edge of the band, the nonlinear mode loses its localization due to coupling with the bulk modes.

One of the most important features of the dislocated Lieb polariton insulator is the robustness of the nonlinear edge states bifurcating from the linear states corresponding to blue and magenta lines in Fig. 2(b). We studied stability of the nonlinear edge states by perturbing them with the broadband input noise (up to 5% in amplitude), so that all pos-sible perturbation modes were excited. We calculated the evolution of these perturbed states over long times using a split-step fast-Fourier method. We found that for nonlinear modes with a positive dispersion $\varepsilon''$ for the associated linear edge state [which is the case for the entire blue and magenta branches in Fig. 2(b)], the perturbed nonlinear edge states evolve as *metastable* objects in the largest part of their existence domain in $\mu$, so that no signs of instability development are seen even at $t=10^3$ (which corresponds to $1.9\,\text{ns}$, notably exceeding the typical lifetime of polariton condensates observed in experiments). An example of robust evolution is shown in Fig. 6 for the edge state corresponding to the red dot in Fig. 3(b) (the energy of this state falls into topological gap, hence coupling with bulk modes is excluded). This is in sharp contrast to the situation encountered in honeycomb lattices [36], where all extended nonlinear Bloch waves are clearly unstable. Metastability was encountered only for nonlinear Bloch modes from the second gap, but not for the modes from the first gap. This phenomenon is one of our central results of this paper, and suggests the use of nonlinear Bloch waves in dislocated Lieb lattices as suitable background for topological dark solitons. Instabilities of Bloch waves from the second gap may show up only very close to the edge of the third band [i.e. close to the dashed line in Fig. 3(b) indicating band edge]. When decreasing the value of $\mu$, the instability development takes more and more time, and in the final run instability becomes undetectable, even though, importantly, the peak amplitude of the corresponding nonlinear edge state is not small (stable solutions can have amplitude $a_- \sim 0.7$), hence nonlinear effects are still strong. At the same time, instabilities are strong for nonlinear states bifurcating from the red and green branches in Fig. 2(b).

The robustness of the topological nonlinear edge states in dislocated Lieb lattice suggests the possibility of existence of topological dark solitons nesting inside the infinitely extended states. To obtain their envelope analytically, we rewrite Eq. (1) as $i\partial \Psi/\partial t = \mathcal{L}\Psi + \mathcal{N}\Psi$, where the operators $\mathcal{L}, \mathcal{N}$ account for the linear and nonlinear terms, respectively, and then substitute the following integral expression $\Psi(x,y,t) = \int_{-K/2}^{+K/2} A(\kappa,t)\mathbf{u}(x,y,k+\kappa)e^{i(k+\kappa)y - i\varepsilon t}d\kappa$. Here the spinor $\mathbf{u} = (u_+, u_-)^{\mathrm{T}}$ solves the linear equation $(\mathcal{L} - \varepsilon)\mathbf{u}e^{iky} = 0$, i.e., it corresponds to the linear Bloch state with momentum $k$. Here, $\kappa$ is the offset from the carrier momentum $k$, and $A(\kappa,t)$ is an unknown spectrally narrow function localized in $\kappa$. Using Taylor series expansion for the spinor $\mathbf{u}(k+\kappa)$ and the energy $\varepsilon(k+\kappa)$, the integral expressions for $\Psi$ and $\mathcal{L}\Psi$ take the form:

$$\Psi(x,y,t) = e^{iky - i\varepsilon t}\sum_{n=0}^{\infty}[i^n n!]^{-1}[\partial^n \mathbf{u}/\partial k^n][\partial^n A(y,t)/\partial y^n], \tag{3}$$

$$\mathcal{L}\Psi(x,y,t) = e^{iky - i\varepsilon t}\sum_{n=0}^{\infty}[i^n n!]^{-1}[\partial^n(\varepsilon \mathbf{u})/\partial k^n][\partial^n A(y,t)/\partial y^n] \tag{4}$$

where $A(y,t) = \int_{-K/2}^{+K/2} A(\kappa,t)e^{i\kappa y}d\kappa$ is the envelope function to be found. Assuming that $\mathbf{u}$ changes with $k$ much slower than the energy $\varepsilon$, one can keep only the $n=0$ term in Eq. (3), so that $\Psi = A\mathbf{u}e^{iky - i\varepsilon t}$, and use $\partial^n(\varepsilon \mathbf{u})/\partial k^n \approx \mathbf{u}\partial^n \varepsilon/\partial k^n$ in Eq. (4), where we keep derivatives of energy $\varepsilon$ up to the second order only. We then project $i\partial\Psi/\partial t = \mathcal{L}\Psi + \mathcal{N}\Psi$ on $\mathbf{u}$ and after some tedious calculations find the required envelope equation:

$$i\frac{\partial A}{\partial t} = -i\varepsilon'\frac{\partial A}{\partial y} - \frac{1}{2}\varepsilon''\frac{\partial^2 A}{\partial y^2} + gA|A|^2, \tag{5}$$

where $\varepsilon'=\partial\varepsilon/\partial k$, $\varepsilon''=\partial^2\varepsilon/\partial k^2$, and $g=\iint\mathbf{u}^\dagger\mathcal{N}\mathbf{u}dxdy/\iint\mathbf{u}^\dagger\mathbf{u}dxdy$ is the positive nonlinear coefficient. When $\varepsilon''>0$ Eq. (5) predicts absence of modulation instability and existence of dark solitons (robust nonlinear edge states are possible exactly in $\varepsilon''>0$ domain). The function describing the envelope of dark soliton is given by $A(y,t)=[(\mu-\varepsilon)/g]^{1/2}\tanh\{[(\mu-\varepsilon)/\varepsilon'']^{1/2}(y-\varepsilon't)\}e^{i(\varepsilon-\mu)t}$, where $\mu-\varepsilon\geq 0$ is the energy shift due to repulsive nonlinearity that should be kept small to ensure that $A(y,t)$ is broad relative to the lattice period. A similar approach to construction of moving bright solitons was applied in [49-52].

In Fig. 7 we show the evolution of a soliton-carrying edge state $\mathbf{\Psi}=A\mathbf{u}e^{iky-i\varepsilon t}$ constructed using the envelope function $A(y,t)$ found above and the Bloch modes $\mathbf{u}=(u_+,u_-)^\mathrm{T}$ from the blue branch of Fig. 2(b) for $\mu-\varepsilon=0.02$ and the momentum $k=0.2\mathrm{K}$ corresponding to $\varepsilon''>0$. The top row shows the nonlinear case, while in the bottom row the nonlinearity in Eq. (1) was switched off. Dark soliton traverses hundreds of lattice periods, but experiences only small oscillations of the width of its notch (see the curve labelled $w_\mathrm{nl}$ in the central panel). No signs of the background instability are visible and there is almost no radiation into the bulk of the lattice. By and large, dark solitons superimposed on metastable nonlinear Bloch wave background are metastable objects too. However, they are excited from rather general input conditions and propagate over long time intervals ($t>10^3$) exceeding the realistic current lifetime of polariton condensates. Therefore, they should be readily observable experimentally. In contrast, the same state strongly disperses in the linear medium, as visible in the evolution of corresponding width $w_{\text{lin}}$ in the central panel of Fig. 7. For the larger energy offsets, $\mu-\varepsilon\sim0.1$, comparable with the width of the existence domain of the nonlinear edge states, the initial dark soliton becomes grey upon evolution and its velocity deviates from $\varepsilon'$.

We also studied the impact of losses on the dynamics of the dark solitons by using a dissipative version of Eq. (1) with the additional loss term $-i\alpha\Psi$ included into the right-hand side of the equation. We used $\alpha=0.01$ corresponding to a polariton lifetime $\sim 190$ ps, as in state-of-the-art experiments [36]. As expected, the soliton amplitude decreases adiabatically and its width self-adjusts in accordance with its instantaneous amplitude. Consistent with expectations, up to $t<1/\alpha$ the dynamics is nonlinear with the width-amplitude ratio closely corresponding to the theoretical predictions. For $t>1/\alpha$ the dynamics becomes effectively linear since the amplitude drops significantly.

Summarizing, we have shown that dislocated Lieb lattice supports topological polariton edge states in both, linear and nonlinear regimes. The nonlinear edge states bifurcating from the linear branches were found to be remarkably robust in contrast to the hexagonal lattices [36]. We also discovered topological dark solitons embedded within the nonlinear edge states. Our results can be applicable for photonic Lieb lattices [53] made from the helical waveguides and the cold atom systems [43,44].

**Funding:** DVS acknowledges support from the ITMO University Visiting Professorship (Grant 074-U01). LT and YVK acknowledge support from the Severo Ochoa (SEV-2015-0522) of the Government of Spain, Fundacio Cellex, Fundació Mir-Puig, Generalitat de Catalunya and CERCA. YVK acknowledges support by the Russian Foundation for Basic Research under the grant 18-502-12080. CL and FY acknowledge support of the NSFC (No. 61475101).

tures of exciton-polaritons in a tunable microcavity with large TE-TM splitting, *Phys. Rev. Lett.* **115**, 246401 (2015).

35. J. Fischer, S. Brodbeck, A. V. Chernenko, I. Lederer, A. Rahimi-Iman, M. Amthor, V. D. Kulakovskii, L. Worschech, M. Kamp, M. Durnev, C. Schneider, A. V. Kavokin, and S. Höfling, Anomalies of a nonequilibrium spinor polariton condensate in a magnetic field, *Phys. Rev. Lett.* **112**, 093902 (2014).

36. Y. V. Kartashov and D. V. Skryabin, Modulational instability and solitary waves in polariton topological insulators, *Optica* **3**, 1228 (2016).

37. D. R. Gulevich, D. Yudin, D. V. Skryabin, I. V. Iorsh, and I. A. Shelykh, Exploring nonlinear topological states of matter with exciton-polaritons: Edge solitons in kagome lattice, *Scientific Reports* **7**, 1780 (2017).

38. F. Baboux, L. Ge, T. Jacqmin, M. Biondi, E. Galopin, A. Lemaître, L. Le Gratiet, I. Sagnes, S. Schmidt, H.E. Türeci, A. Amo, J. Bloch, Bosonic condensation and disorder-induced localization in a flat band, *Phys. Rev. Lett.* **116**, 066402 (2016).

39. S. D. Huber, E. Altman, Bose condensation in flat bands, *Phys. Rev. B* **82**, 184502 (2010).

40. M. Biondi, E.P.L. van Nieuwenburg, G. Blatter, S.D. Huber, S. Schmidt, Incompressible polaritons in a flat band, *Phys. Rev. Lett.* **115**, 143601 (2015).

41. C. E. Whittaker, E. Cancellieri, P. M. Walker, D. R. Gulevich, H. Schomerus, D. Vaitiekus, B. Royall, D. M. Whittaker, E. Clarke, I. V. Iorsh, I. A. Shelykh, M. S. Skolnick, D. N. Krizhanovskii, "Exciton-polaritons in a two-dimensional Lieb lattice with spin-orbit coupling, arXiv:1705.03006 (2017).
18

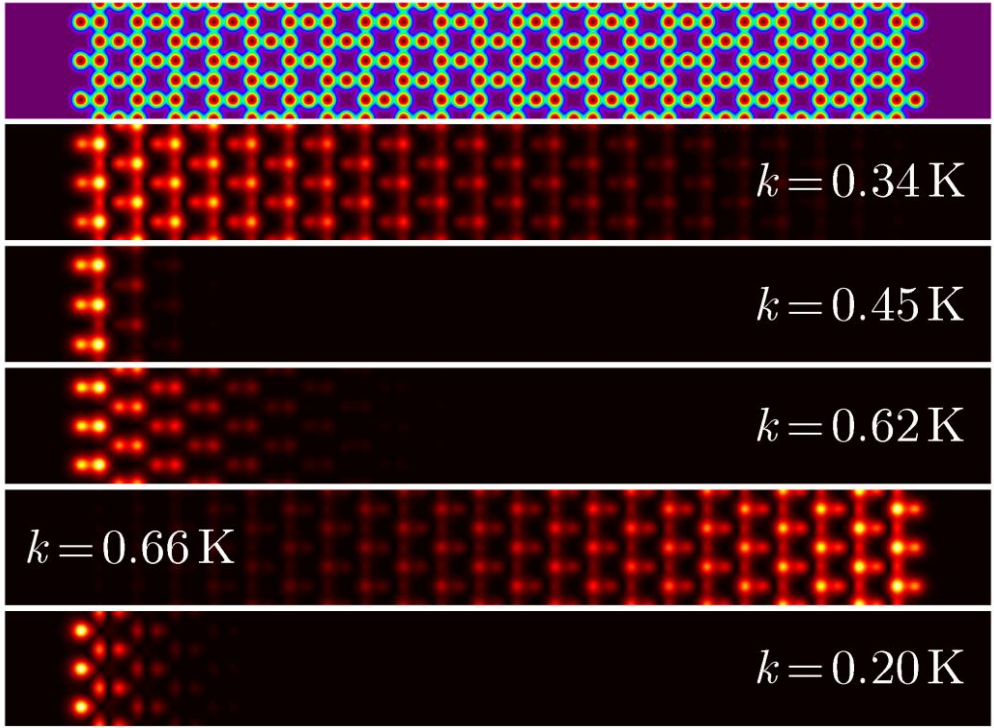

Figure 1. Dislocated Lieb lattice (first row) and examples of linear edge states associated with green (second-forth rows), red (fifth row), and blue (sixth row) branches in the eigenvalue spectrum depicted in Fig. 2(b). Only the dominating $|\psi_-|$ spinor component is shown. In all cases $\beta=0.3$, $\Omega=0.5$.

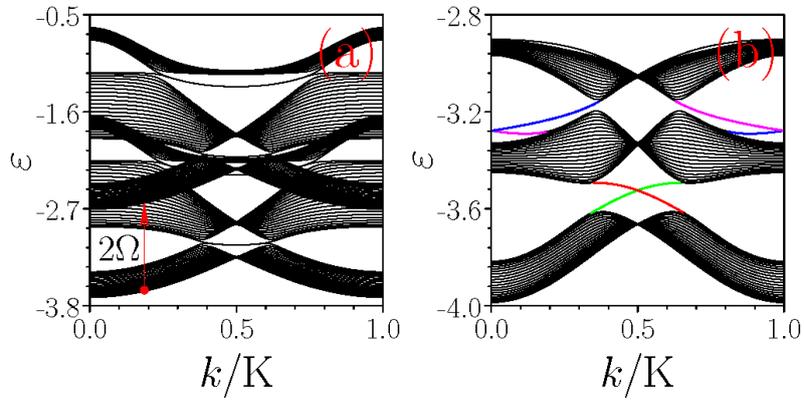

Figure 2. Energy-momentum diagrams $\varepsilon(k)$ for a truncated dislocated Lieb lattice for $\beta = 0$ (a), $\beta = 0.3$ (b) and $\Omega = 0.5$. Red, green, blue and magenta lines correspond to unidirectional topological edge states, while black lines correspond to bulk modes.

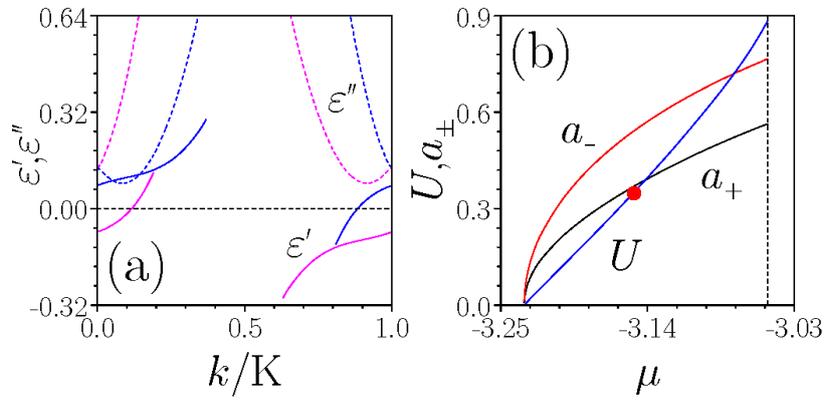

Figure 3. Properties of the linear and nonlinear edge states. (a) First-order $\varepsilon'$ and second-order $\varepsilon''$ derivatives of the energy of the linear edge state versus momentum $k$. (b) Peak amplitudes of $\psi_\pm$ components and norm per $y$-period at $k=0.20\,\mathrm{K}$ versus $\mu$ for the nonlinear edge states. Dashed line indicates the border of the topological gap. In all cases $\beta=0.3$, $\Omega=0.5$.

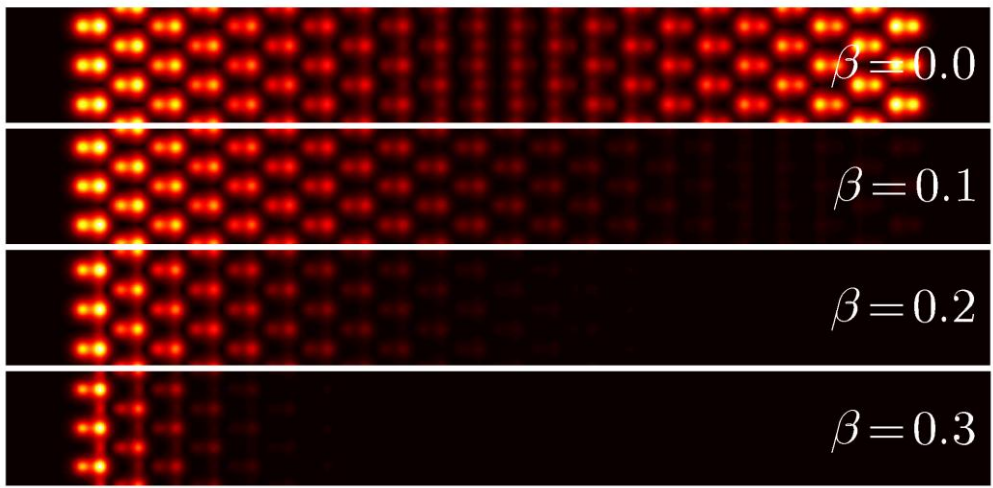

Figure 4. Transformation from a non-topological to a topological edge state (green branch from Fig. 2) with increase of $\beta$, at $k = 0.40\,\text{K}$. Only $|\psi_-|$ is shown.

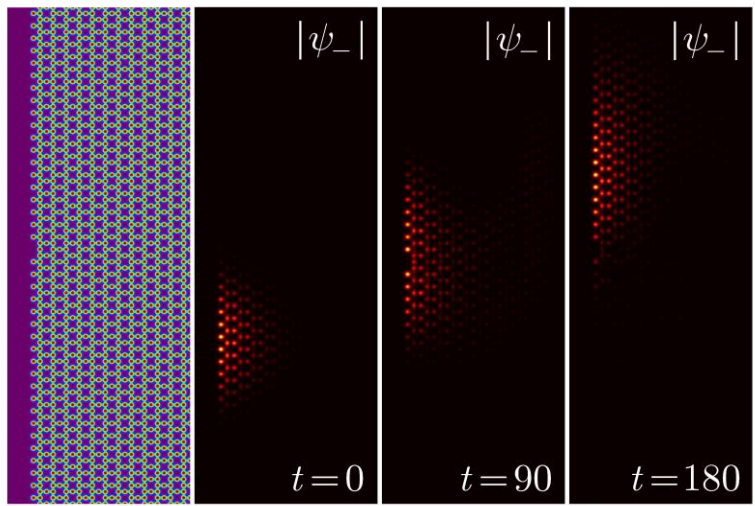

Figure 5. Passage of linear edge state with $k = 0.30\,\mathrm{K}$ and broad envelope from blue branch in Fig. 2(b) through surface defect without noticeable backscattering at $\beta = 0.3$, $\Omega = 0.5$.

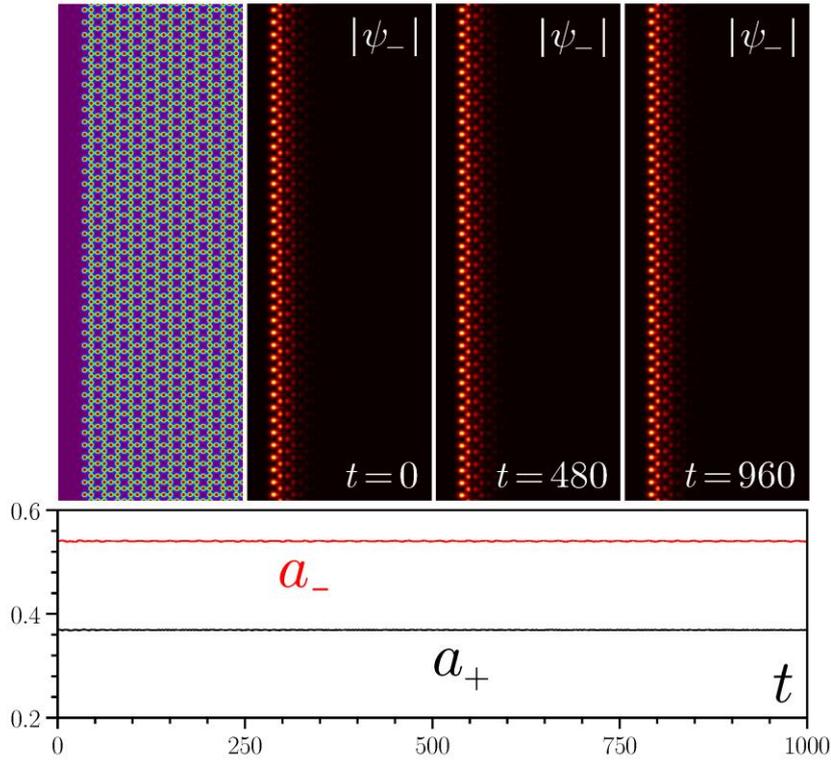

Figure 6. Robust evolution of the perturbed nonlinear edge state from blue branch corresponding to $\mu=-3.15$, $k=0.20\,\mathrm{K}$, $\beta=0.3$, $\Omega=0.5$ [red dot in Fig. 3(b)]. Top: lattice profile (40 y-periods) and $|\psi_-|$ distributions at different moments of time. Bottom: evolution of peak amplitudes of components.

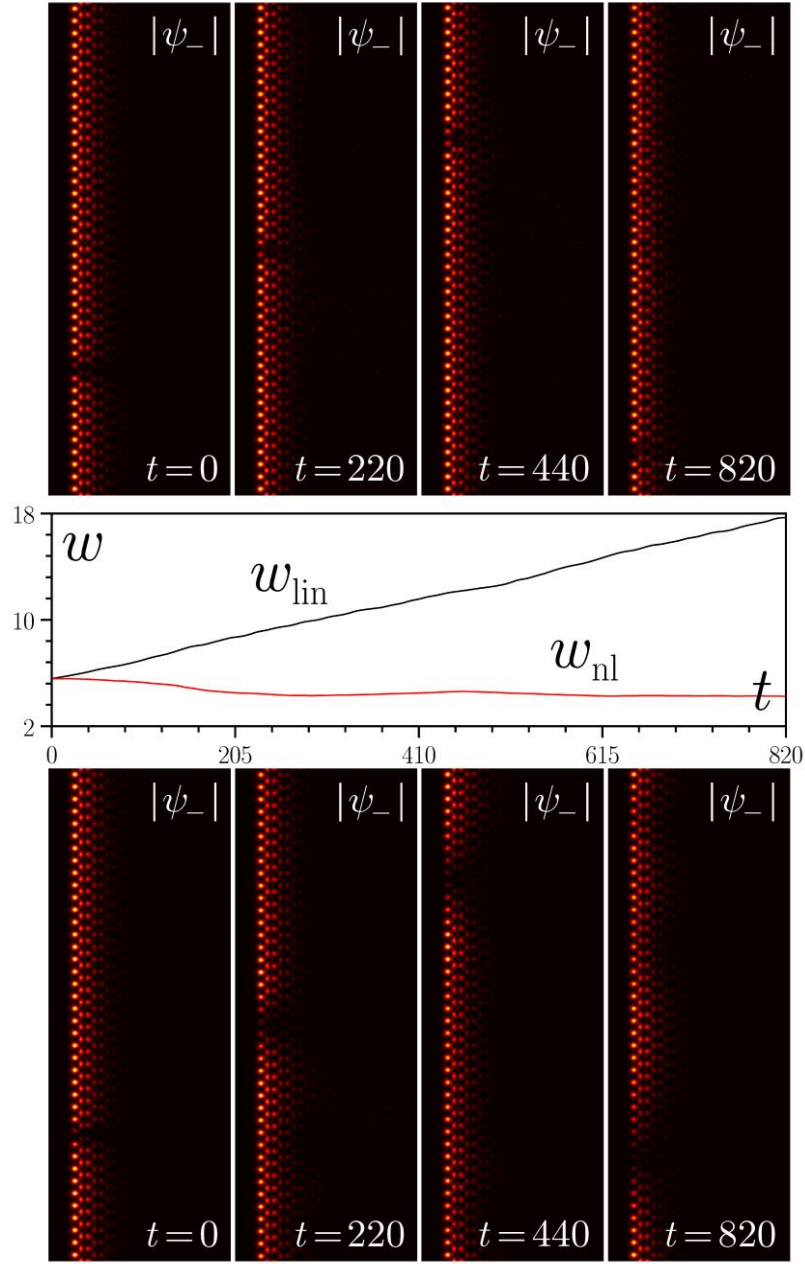

Figure 7. Stable evolution of a dark soliton nested in the edge state (blue branch in Fig. 2) at $\mu-\varepsilon=0.02$, $k=0.20\,\mathrm{K}$, $\beta=0.3$, $\Omega=0.5$ in the nonlinear case (top row) and its spreading in the linear case (bottom row). The middle row shows the width of the dark spot as a function of time. Due to strong vertical displacement experienced by soliton, the distribution at $t=820$ was shifted vertically.

27